# Optical spin readout of a silicon color center in the telecom L-band

Shuyu Wen,[1,2,3] Gregor Pieplow,[4] Junchun Yang,[1,5] Kambiz Jamshidi,[5] Manfred Helm,[1,5] Jun-Wei Luo,[2,3] Tim Schröder,[4] Shengqiang Zhou,[1*] & Yonder Berencén[1**]

[1]*Helmholtz-Zentrum Dresden-Rossendorf, Institute of Ion Beam Physics and Materials Research, Bautzner Landstrasse 400, 01328 Dresden, Germany*

[2]*State Key Laboratory of Superlattices and Microstructures, Institute of Semiconductors, Chinese Academy of Sciences, Beijing 100083, China*

[3]*Center of Materials Science and Optoelectronics Engineering, University of Chinese Academy of Sciences, Beijing 100049, China*

[4]*Humboldt-Universität zu Berlin, Department of Physics 12489 Berlin, Germany*

[5]*Technische Universität Dresden, 01062 Dresden, Germany*

Corresponding authors: s.zhou@hzdr.de[*]; y.berencen@hzdr.de[**]

**Silicon-based quantum technologies have gained increasing attention due to their potential for large-scale photonic integration, long spin coherence times, and compatibility with CMOS fabrication. Efficient spin-photon interfaces are crucial for quantum networks, enabling entanglement distribution and information transfer over long distances. While several optically active quantum emitters in silicon have been investigated, no spin-active defect with optical transitions in the telecom L-band (1565–1625 nm)—a key wavelength range for low-loss fiber-based communication—has been experimentally demonstrated. Here, we demonstrate the optical detection of spin states in the C center, a carbon-oxygen defect in silicon that exhibits a zero-phonon line at 1571 nm. By combining optical excitation with microwave driving, we achieve optically**

**detected magnetic resonance (ODMR), enabling spin-state readout via telecom-band optical transitions. These findings provide experimental validation of recent theoretical predictions and mark a significant step toward integrating spin-based quantum functionalities into silicon photonic platforms, paving the way for scalable quantum communication and memory applications in the telecom L-band.**

**Introduction**

Silicon has emerged as a leading platform for photonic quantum technologies, driven by its compatibility with CMOS fabrication processes, long spin coherence times, and its potential for scalable photonic integration[1–3]. Among the various quantum systems being explored in silicon, optically active spin-based qubits have gained significant interest in quantum communication, sensing, and computation[4]. A key challenge in these areas is the realization of efficient spin-photon interfaces, which are essential for quantum state transfer, entanglement distribution, and long-distance quantum communication[4,5].

Despite significant progress in identifying single color centers in silicon, the realization of a spin-photon interface operating in the telecom L-band (1565–1625 nm) remains an open challenge. Several single-color centers, including the W, G, and T centers, have been explored for photonic quantum technologies[6–8]. Among them, the T center is the only silicon defect that has demonstrated an optically active spin-photon interface, exhibiting coherent spin properties and optical transitions in the O-band (1260–1360 nm)[7]. This makes it a promising candidate for scalable quantum memory and distributed quantum computing, as demonstrated by recent remote entanglement experiments[9]. However, while the O-band is advantageous for silicon photonics integration, it is less optimal for long-distance fiber-based quantum communication due to higher attenuation in silica fibers compared to the L-band.

The C center, a carbon-oxygen defect in silicon[10], emits in the telecom L-band, making it a promising contender for an L-band spin-photon interface. Historically studied for its photoluminescence properties[11–16], its spin properties have only recently been theoretically predicted[17]. Unlike the T center, which has a doublet ground state[7], the C center possesses a singlet ground state[17], effectively isolating nearby nuclear spin qubits from magnetic perturbations. This latter enhances nuclear spin coherence, potentially extending nuclear memory lifetimes, making the C center particularly attractive for long-lived quantum memories with potential optical spin readout and control.

In this work, we experimentally investigate the C center in silicon, demonstrating optical access to its spin states. Using silicon-on-insulator (SOI) substrates, we fabricate and characterize these defects, achieving optically detected magnetic resonance (ODMR) through combined microwave (MW) and optical excitation. By probing the spin-triplet (S = 1) excited state and its transitions from singlet (S = 0) states, we establish the C center as a viable spin-photon interface in silicon. Our results provide insights into optical spin readout, marking a critical step toward the realization of long-lived quantum memory and the development of scalable spin-photonic technologies in silicon.

**Results**

Carbon and oxygen are common impurities in silicon introduced during wafer fabrication[18], which can lead to the formation of various color centers when exposed to radiation[14]. As illustrated in Fig. 1 a, the C center is one such carbon-related color center, formed by a carbon-oxygen interstitial pair ($C_i$-$O_i$) within the silicon lattice. The $C_i$-$O_i$ pairs can capture a hole through short-range defect potential and subsequently capture an electron via long-range Coulomb potential, which facilitates the formation of a hole-attractive isoelectronic bound exciton (IBE)[11,12]. This process gives rise to a series of excited state energy levels ($C_i$ with i=T, 0, 1,…,9)[13].

First, we investigate the optical properties of an ensemble of C centers through PL measurements at 20 K, as shown in Fig. 1 b. Excitation can be achieved either near (1.16 eV, 1064 nm) or below (0.95 eV, 1310 nm) the silicon bandgap, with the latter enabling excitation within the 1310 nm telecom O-band while collecting PL emission in the telecom C- and L-bands. The most prominent emission peak, corresponding to the $C_0$ line, appears at 1571 nm in the telecom L-band and is attributed to the zero-phonon line (ZPL) transition from the $C_0$ excited state to the ground state. A second emission peak, identified as the $C_1$ line at 1560 nm in the telecom C-band, originates from the ZPL transition from the $C_1$ excited state to the ground state. Additionally, a broad peak around 1608 nm is observed, corresponding to the $C_0$ phonon sideband ($C_0^{TA}$).

Further temperature-dependent and laser power-dependent PL results are presented in the supplementary material (Fig. S1). The temperature-dependent PL measurements reveal significant thermal quenching of the PL signal as the sample temperature increases, with the C center PL emission lines being fully quenched above 45 K. This behavior is consistent with the excitonic properties of the C center. The PL quenching at higher temperatures likely results from a thermally activated reverse intersystem crossing process that equalizes the excited states[17].

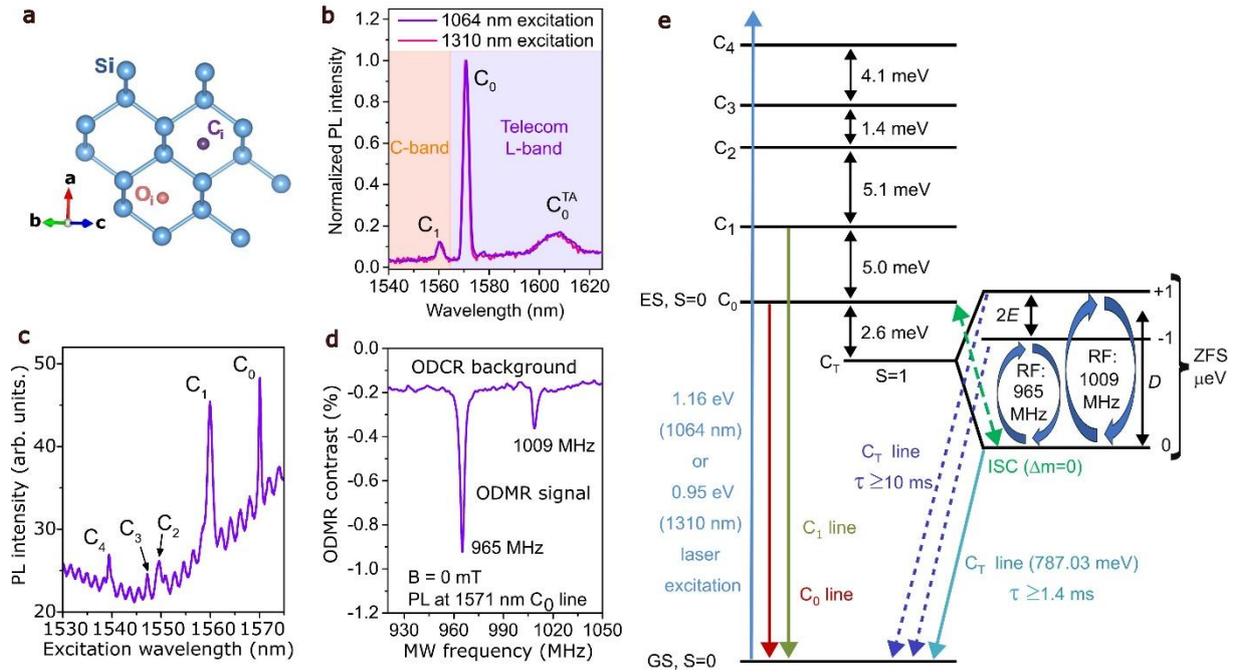

**Figure 1 | Optical and spin properties of the C center in silicon.**

**a**, Atomic structure of the C center in silicon, consisting of an interstitial oxygen atom ($O_i$) and an interstitial carbon atom ($C_i$), which together form the $C_i$-$O_i$ pair within the silicon host. **b**, Photoluminescence (PL) spectrum of an ensemble of C centers at 20 K, excited using a telecom O-band (1310 nm) laser, with PL emission collected from both the telecom C-band ($C_1$ line) and L-band ($C_0$ line). **c**, Photoluminescence excitation (PLE) measurements performed by sweeping the excitation wavelength of a tunable laser and collecting the signal from the $C_0$ phonon sideband ($C_0^{TA}$), using a 1610 nm bandpass filter with an FWHM bandwidth of 12 nm. **d**, Optically detected magnetic resonance (ODMR) measurements under 1064 nm laser excitation showing two resonant peaks under zero external magnetic field, with the PL signal spectrally collected only from the 1571 nm $C_0$ line. The ODMR signal is superimposed on an optically detected cyclotron resonance (ODCR) background. **e**, Energy level diagram and optical transitions of the C center in silicon. The diagram illustrates the key energy levels of the C center, including the $C_T$ state and the higher-lying excited states ($C_0$, $C_1$, $C_2$, $C_3$, and $C_4$), as well as the corresponding optical transitions between these states and the ground state. The $C_T$ state is optically active at temperatures around 2 K[16]. The energy levels are not drawn to scale.

To probe the energy level structure of the C center, we conducted photoluminescence excitation (PLE) measurements by sweeping the excitation wavelength of a tunable laser and detecting photoluminescence from the $C_0$ phonon sideband ($C_0^{TA}$) at 1610 nm. The resulting PLE spectrum (Fig. 1 c) displays a series of resonant peaks corresponding to the $C_0$ (1571 nm), $C_1$ (1560 nm), $C_2$ (1549 nm), $C_3$ (1547 nm), and $C_4$ (1539 nm) states. These peaks, which are separated by only a few meV, indicate distinct excited states within the C center, which arise from the excitonic nature of the center itself. Our observations suggest carrier transport from the higher-lying states ($C_1$-$C_4$) to the lower-lying $C_0$ state. Background oscillations in the PLE spectrum are attributed to Fabry-Perot resonances originating from the 12 µm-thick SOI device layer.

The ODMR properties of the C center in silicon were recently predicted by Udvarhelyi et al[17]. Theoretical simulations indicate that the $C_T$ energy level, characterized by S=1, exhibits a zero-field splitting (ZFS) between the $m_s$=0 and $m_s$=±1 states. Due to the non-radiative properties of the $C_T$ line at 20 K and the significant lifetime difference between the $m_s$=0 state (> 1.4 ms) and the $m_s$=±1 states (> 10 ms), resonant spin pumping via radio frequency (RF) signals can promote carriers from the $m_s$=0 state to the $m_s$=±1 states, resulting in a prolonged non-radiative decay time. This extended non-radiative process can trap a portion of the carriers, reducing the number of carriers involved in optical transitions between the excited states ($C_0$, $C_1$) and the ground state, leading to decreased PL intensity during RF resonant pumping.

While ODMR measurement and spin coherent control have been experimentally achieved in wide-bandgap materials such as diamond[19], silicon carbide (SiC)[20], and hexagonal boron nitride (hBN)[21], achieving ODMR on a silicon platform poses significant challenges. The smaller bandgap of silicon allows for strong pumping of photoexcited free carriers by the microwave electric field[22]. Additionally, non-resonant RF-induced heating effects can further complicate the situation, resulting in a strong RF-independent background due to optically detected cyclotron resonance (ODCR) effects, which obscures the RF-dependent ODMR signal[23].

To overcome these challenges, we combine optical excitation, either near (1.16 eV, 1064 nm) or below (0.95 eV, 1310 nm) the silicon bandgap, with a pulsed RF excitation scheme using a coplanar RF waveguide. This approach helps minimize the ODCR background by mitigating the free carrier heating effect, thereby achieving a clear ODMR signal from the C centers.

The ODMR measurement results for an ensemble of C centers under 1064 nm optical excitation and without an external magnetic field are presented in Fig. 1 d. This data reveals a typical zero-field splitting (ZFS) characterized by two sharp resonant peaks centered at 987 MHz, indicative of an effective longitudinal ZFS parameter $D/h$=987 MHz. The main ODMR resonant signal consists of these two sharp peaks located at 965 MHz and 1009 MHz, which represent the E-splitting between the $m_s$=+1 and $m_s$=-1 $C_T$ sublevels, yielding an effective transverse ZFS parameter $E/h$= 22 MHz.

We also measured $\Delta$PL=$PL_{RF-ON}$-$PL_{RF-OFF}$ at various PL emission wavelengths, as shown in Fig. S2. Our PL measurements reveal that only the $C_0$ and $C_1$ PL emission lines are detectable under the given conditions, with both energy levels suitable for ODMR readout. The optical transition from the $C_1$ energy level to the ground state ($C_1$ line at 1560 nm) exhibits ODMR at a resonant MW frequency of 965 MHz. This suggests that both $C_0$ and $C_1$ energy levels relax via the $C_T$ state, with the $C_0$ state potentially involved through an intersystem crossing (ISC) transition. Alternatively, carrier transfer between the $C_0$ and $C_1$ energy levels could explain the observed behavior. This phenomenon has not been predicted in Ref. [17], warranting further investigation.

Based on the optical spectroscopy and ODMR measurements presented in Fig. 1 b-d, we propose the energy level diagram of the C center in silicon, as shown in Fig. 1 e. While the structure closely resembles the theoretical model described in Ref. [17], we identify differences in the transition energies and observe that $C_1$ is also coupled to the spin-triplet state ($C_T$).

We thus propose the following mechanism for the ODMR of the C center based on our experimental observations: In the optical pumping process without resonant RF excitation, the majority of carriers are directly excited from the ground state to the higher-lying excited states (blue solid line in Fig. 1 e), subsequently transitioning back to the ground state on a timescale of tens of microseconds via radiative transitions with photon emission at 1571 nm (red solid line in Fig. 1 e). A small fraction of carriers undergoes a transition from the $C_0$ level to the $m_s=0$ sublevel of the $C_T$ triplet energy level via the ISC (green dashed line in Fig. 2), according to the spin selection rule $\Delta m_s=0$. These carriers then experience a non-radiative decay process to the ground state, with a decay time exceeding 1.4 ms (aqua blue solid line in Fig. 1 e).

When the RF matches the energy gap between the $m_s=0$ and $m_s=-1$ spin sublevels (blue curved arrows, 965 MHz, in Fig. 1 e) or between the $m_s=0$ and $m_s=+1$ spin sublevels (blue curved arrows, 1009 MHz, in Fig. 1 e) of the $C_T$ energy level, carriers in the $m_s=0$ spin sublevel can be resonantly pumped to the $m_s=\pm1$ spin sublevels. The $C_T$ level, located 2.64 meV below the $C_0$ level, exhibits a significantly longer decay lifetime to the ground state, making it a dark energy level in the ODMR experiments. This transition leads to an extended non-radiative decay time (>10 ms) from the $m_s=\pm1$ sublevels to the ground state (dark blue dash lines in Fig. 1 e), compared to the shorter decay time (>1.4 ms) from the $m_s=0$ sublevel to the ground state. Under RF resonant pumping, this prolonged non-radiative decay process results in carrier trapping within the long-lifetime $m_s=\pm 1$ spin sublevels, reducing the carrier population available for optical transitions involving $C_0$ and $C_1$ (shown by the red and green solid lines in Fig. 1 e). Consequently, this leads to a reduction in PL intensity during RF resonant pumping in the $C_T$ energy level, resulting in a negative ODMR contrast as experimentally observed in Fig 1 d.

The observation of the ODMR signal represents the first-ever demonstration of a spin-optical interface for the C center, achieved by RF resonant pumping between the $m_s=0$ and $m_s=\pm 1$ spin sublevels of the $C_T$ triplet energy level. To optically read out the spin-pumping effect, one can detect the PL differences during the optical transitions from the $C_0$ and $C_1$ energy levels to the ground state. This capability enables spin-readout in the telecom C- and L-bands.

To further investigate the ODMR signal and ODCR background, we conducted temperature-dependent, laser-power-dependent, and RF-power-dependent ODMR measurements on an ensemble of C centers. The results of the temperature-dependent ODMR measurements, shown in Fig. 2 a, indicate that both the ODMR signal and the ODCR background increase as the temperature rises from 20 K to 25 K. However, the ODCR background exhibits a significantly stronger increase and largely overlaps with the ODMR signal. This substantial rise in the ODCR background may be attributed to electron cyclotron resonance excited by the MW electric field, which leads to enhanced free-carrier heating effects at elevated lattice temperatures[22,23].

The laser-power-dependent ODMR measurements, depicted in Fig. 2 b, reveal that the ODCR background remains constant even as the excitation laser power decreases by an order of magnitude. This finding suggests that the ODCR background is not directly related to the heating effect of the laser. In contrast, the RF-power-dependent ODMR measurements presented in Fig. 2 c demonstrate that the ODCR background is highly sensitive to the RF excitation power, further confirming its origin in the RF-driven electron cyclotron resonance effect.

The trends of both the ODMR signal and the ODCR background are illustrated in Fig. 2 d. Notably, the ODCR background exhibits a saturation tendency when the RF power exceeds 100 mW, whereas no similar saturation tendency is observed in the ODMR signal.

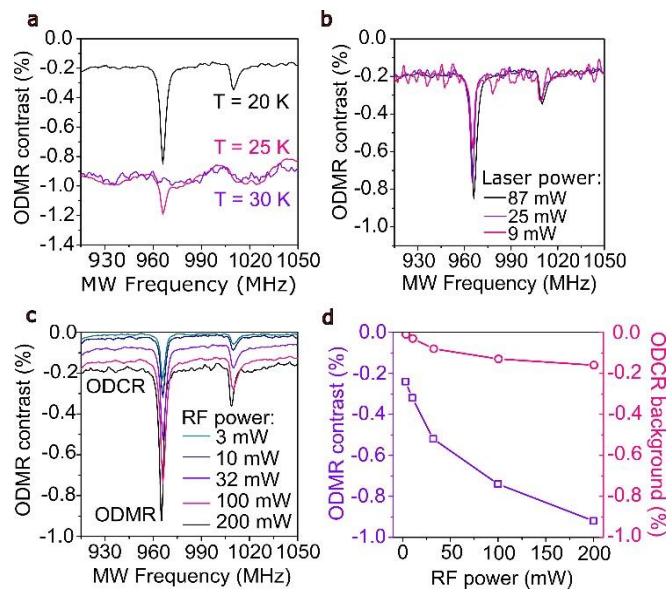

**Figure 2 | Dependence of ODMR contrast on temperature, laser power, and RF power.**

**a**, ODMR contrast as a function of RF at different temperatures (20 K, 25 K, and 30 K), showing a reduction in contrast with increasing temperature. **b**, ODMR contrast versus RF measured at varying laser powers (9 mW, 25 mW, and 87 mW), indicating minimal dependence on laser power. **c**, ODMR contrast as a function of RF measured at different RF powers, illustrating an increase in ODCR background with increasing RF power. **d,** ODMR contrast (purple squares) and ODCR background signal (pink circles) as a function of RF power, demonstrating an increasing ODCR contribution at higher RF power.

The $C_T$ spin triplet is a crucial energy level that facilitates the spin-optical interface in ODMR. The Hamiltonian (*H*) governing a single C-center in the presence of a magnetic field ***B*** is expressed as[12]

$$H = H_0 + H_{SS} + H_{SO} + H_Z, \qquad (1)$$

where $H_0$ represents the unperturbed Hamiltonian, $H_{SS}$ the spin-spin (SS) interaction, $H_{SO}$ the spin-orbit (SO) coupling and $H_Z$ the Zeeman interaction. The spin-spin term is given by

$$H_Z = D'(S_z^2 - S(S+1)/3) + E'(S_+^2 + S_-^2), \qquad (2)$$

where $S=1$, $S_\pm = S_x \pm iS_y$, $S_{x,y,z}$ denote components of the spin operator (***S***), $D'$ and $E'$ correspond to the longitudinal and transverse zero-field splitting (ZFS) parameters, respectively. The spin-orbit interaction is described as

$$H_{SO} = \lambda_z L_z S_z + \lambda_+ L_+ S_+ + \lambda_- L_- S_-, \qquad (3)$$

where $L_\pm = L_x \pm iL_y$, $L_{x,y,z}$ denote the components of the angular momentum operator (***L***) and $\lambda_{z,\pm}$ are the spin-orbit coupling constants. According to Udvarhelyi et al.[17], the ordering of the singlet and triplet states at ***B***=0, as shown in Fig. 1 e, arises from both the spin-spin and spin-orbit interaction terms. These interactions also determine the effective transverse zero-field splitting, *D/h*=987 MHz, and the longitudinal splitting, *E/h*=22 MHz, as illustrated in Fig. 1 e. However, these measured values of *D/h* and *E/h* deviate from the theoretical predictions of Udvarhelyi et al.[17].

This discrepancy may be attributed to the omission of vibrational degrees of freedom in their model, which can suppress spin-orbit coupling[24].

The Zeeman interaction is expressed as:

$$H_Z = \mu_b \boldsymbol{B} \cdot (\mathbf{L} + g_0 \mathbf{S}), \qquad (4)$$

where $\mu_b$ is the Bohr magneton and $g_0$ is the Landé g-factor. This interaction can be simplified to $H_Z = \mu_b \boldsymbol{B} \hat{g} \mathbf{L}$, where $\hat{g} = \text{diag}\{g_\perp, g_\perp, g_\parallel\}$. The Zeeman term splits $C_T$ energy levels in response to the strength and orientation of an external magnetic field, inducing shifts in the ODMR frequency. The anisotropic nature of the $\hat{g}$-tensor reflects the orbital-Zeeman contribution to triplet-state splitting. Overall, the $H_Z$ term provides sensitivity to external magnetic fields, which is central to the application of the C center as a spin-optical interface.

To investigate the spin properties of the C center, we measure its ODMR response under external magnetic fields with varying strengths and orientations. Figure 3 shows the ODMR splitting of the C center as a function of the external magnetic field direction and strength. Figure 3a schematically illustrates the measurement setup, where ODMR is measured for three different orientations of the external magnetic field ($B_x$, $B_y$, $B_z$). Here, in-plane magnetic fields $B_x$ and $B_y$ are aligned parallel to the silicon (110) crystal plane in the device layer of the SOI sample, with $B_z$ perpendicularly aligned to the [110] crystal orientation of silicon.

Figures 3 b-d show the ODMR signal's splitting behavior as the external magnetic field strength increases. This data reveals a strong anisotropy in the ODMR splitting of the C center, resembling the behavior observed in ensembles of NV⁻ centers in diamond[25], which share a similar crystal structure to silicon. The pronounced anisotropy of ODMR splitting suggests that the C center exhibits multiple quantization axes within the silicon lattice. When the external magnetic field is misaligned with a quantization axis, the Zeeman splitting of the $C_T$ energy level becomes highly sensitive to the misalignment angle, resulting in a complex pattern of ODMR splitting components in C center ensembles (Fig. 3 c and d).

In Figure 3 b, when the external magnetic field is applied along the $B_y$ axis, the number of splitting peaks is minimized to four, indicating alignment with one of the high-symmetry spin quantization axes, similar to the ensemble behavior seen in NV⁻ centers in diamond[25]. However, as the misalignment between the quantization axis and the external magnetic field increases (as seen from Figure 3 c to Figure 3 d), the number of ODMR splitting peaks rises significantly, in good agreement with the expected spin Hamiltonian behavior[25]. These results suggest that the highly anisotropic ODMR splitting and the complexity observed in C-center ensembles originate from the tensorial nature of $\hat{g}$ and the presence of multiple quantization axes. Notably, the analysis of ODMR spectra could be significantly simplified when scaling down to a single defect[26].

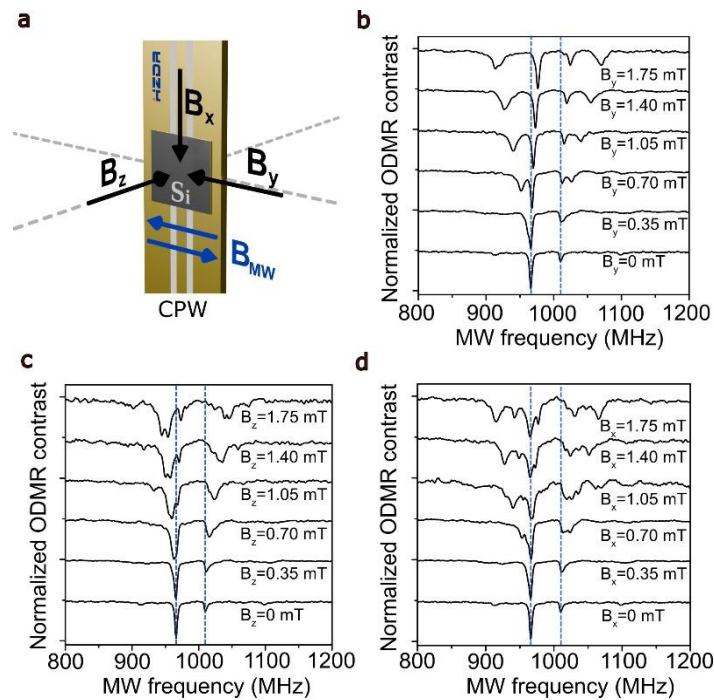

**Figure 3 | Anisotropic ODMR splitting of the C center in silicon.**

**a**, Schematic of the ODMR measurement setup, illustrating the magnetic field orientations applied to the silicon sample containing C centers. The external magnetic field components are defined as follows: $B_x$ — in-plane magnetic field along the X-axis, parallel to the coplanar waveguide (CPW); $B_y$ — in-plane magnetic field along the Y-axis; and $B_z$ — out-of-plane magnetic field along the Z-axis. (b-d) Optically detected magnetic resonance (ODMR) spectra measured under different external magnetic field orientations: (b) $B_y$, (c) $B_z$, and (d) $B_x$. The ODMR spectra exhibit anisotropic splitting as the external magnetic field increases. The Y-axis values are offset for clarity.

**Discussion**

Through ODMR spectroscopy, we have identified a spin-state manifold in the C center that couples to optical transitions in the telecom L-band, revealing a level structure reminiscent of the TR12 defect in diamond[27]. This structure consists of singlet ground and excited states, along with an intermediate triplet excited state. Our observations align with recent DFT predictions[17] and provide more precise quantification of the optical and RF transition energies.

While coherence studies and the isolation of a single C center are still in progress, its lack of dark spin sublevels implies a potentially more stable spin state, offering longer coherence times. This is a key feature for applications in quantum memory and entanglement distribution, where the stability and fidelity of spin states are crucial.

Optimizing the C center for quantum applications requires careful consideration of the isotopic composition of the surrounding silicon, carbon, and oxygen atoms. Natural silicon contains approximately 5% of the isotope $^{29}$Si, which has a nuclear spin ½ that can interact with the C center's spin, potentially limiting coherence times[28]. Therefore, isotopically purified $^{28}$Si substrates are preferred to reduce these unwanted interactions and achieve more stable spin states. Additionally, while $^{16}$O does not have a nuclear spin, the presence of oxygen isotopes could still influence the spin dynamics of the C center, particularly through hyperfine interactions with the carbon nucleus[29].

The natural abundance of $^{13}$C in silicon is approximately 1%, and while it also possesses a nuclear spin ½, its presence can be beneficial for exploring quantum memory applications. The nuclear spin of $^{13}$C nuclei may provide an additional resource for nuclear spin-based information processing, as shown in previous studies on similar systems, such as isotopically engineered diamond[30]. While isotopic purification of $^{12}$C and $^{16}$O reduces unwanted spin interactions, the inclusion of $^{13}$C introduces the potential for more complex quantum information processing, including nuclear spin-based quantum memory[31]. In particular, for a single C center, the $C_T$ triplet excited state can be employed to initialize and read out nearby $^{13}$C nuclear spin qubits. Unlike the T center, which has a doublet ground state[7], the C center's ground state is a singlet. This singlet nature effectively isolates nearby nuclear spin qubits from magnetic perturbations when the C center is in its ground state, thereby preserving nuclear spin coherence and potentially extending nuclear memory lifetimes[32]. Moreover, the C center's ZPL, which will need to be enhanced through coupling to an optical cavity for Purcell enhancement, might offer a robust interface between stationary nuclear spin qubits and flying qubits (single photons)[4].

A key requirement for developing a spin-photon interface that exploits the coupling between spin sublevels and the optical ZPL in the L-band is the establishment of coherent interactions between the singlet and triplet states. Demonstrating such coherence via intersystem crossing using coherent quantum control remains an open challenge and would represent a significant advancement for solid-state spin defects.

However, this capability is essential for more advanced quantum technologies, including quantum communication and information processing.

In summary, our results establish the C center in silicon as a promising spin-optical interface in the telecom L-band. The potential to coherently couple spin states to optical transitions in this wavelength range marks a significant step toward spin-based quantum technologies, with applications in quantum memory, entanglement distribution, and secure communication. Enhancing the C center's environment—through isotopically purified $^{28}$Si substrates[33] and controlled carbon-oxygen compositions—could improve spin coherence and mitigate unwanted interactions. Additionally, the presence of $^{13}$C offers further opportunities for nuclear spin-based quantum memories, advancing quantum information processing.

**Methods**

*Creation of the C center*

We implanted carbon into the [110]-oriented, 12 µm-thick device layer of silicon-on-insulator substrates at a fluence of 4×10$^{13}$ cm$^{-2}$ and an energy of 30 keV. Following implantation, the samples underwent annealing at 1000°C for 20 seconds in a nitrogen atmosphere to repair lattice damage and incorporate the carbon atoms into substitutional sites within the silicon lattice. After annealing, the samples were subjected to proton irradiation at a fluence of 4×10$^{13}$ cm$^{-2}$ at an energy of 1 MeV.

*PL measurement*

We conducted optical excitation using either a 1064 nm or a 1310 nm laser (Thorlabs, Inc.), isolating the photoluminescence signal from the laser light with a 1500 nm dichroic mirror and a 1450 nm long-pass filter. The PL spectrum was collected with a TRIAX 550 monochromator paired with InGaAs photodetectors (IGA-020-E-LN7). To synchronize the laser signal, we used an optical chopper (Stanford Research SR540), and a lock-in amplifier (Stanford Research SR830) was employed to detect the PL signal. The 5 mm×5 mm silicon-on-insulator sample containing C centers was mounted in a closed-cycle helium cryostat and cooled to 20 K.

*PLE measurement*

We performed optical excitation using a tunable laser (TL1550-B, Thorlabs, Inc.) in combination with an Aerodiode SOA-4-3-2 semiconductor optical amplifier. The resulting PL signal was collected through a 50:50 beam splitter, followed by a 1610 nm bandpass filter (FWHM bandwidth = 12 nm) for spectral isolation. The integrated PL signal was then detected by InGaAs photodetectors (IGA-020-E-LN7) connected to a lock-in amplifier system (Stanford Research SR830) for precise signal measurement.

*ODMR measurement*

We performed optical excitation using either a 1064 nm or 1310 nm laser (Thorlabs, Inc.). To isolate the photoluminescence signal from laser excitation, we employed a 1500 nm dichroic mirror followed by a long-pass filter at 1450 nm. The PL signals at different wavelengths ($C_0$ line at 1571 nm and $C_1$ line at 1560 nm) were collected through a TRIAX 550 monochromator combined with InGaAs photodetectors (IGA-020-E-LN7).

For precise PL measurement, we employed synchronization and lock-in detection techniques. A 22 Hz synchronization signal, generated internally by a lock-in amplifier (Stanford Research SR830), was used to synchronize the entire setup. The lock-in amplifier amplitude-modulated (AM) the RF output from the Anapico ASPIN6G RF signal generator with this 22 Hz signal, creating a periodic RF-on/RF-off cycle. During these synchronized RF-on/off states, the lock-in amplifier detected the difference in PL (ΔPL) with and without RF excitation to obtain the ΔPL readout.

Since the ΔPL signal amplitude was significantly lower than the overall PL level, we connected an electronic filter (Thorlabs, EF500) with a 1 dB cutoff of >1 Hz between the InGaAs photodetector and the lock-in amplifier. This filter blocked the direct component of the output current, enabling the use of a higher-sensitivity circuit in the lock-in amplifier, thereby enhancing the signal-to-noise ratio for more accurate measurements.

ODMR signal versus the RF was then recorded by sweeping the RF and detecting changes in the PL intensity of the $C_0$ line (1571 nm) with and without RF excitation. The ODMR contrast was deduced as follows:

$$ODMR\ contrast = \frac{\Delta PL}{PL} \times 100\% = \frac{PL_{RF-ON} - PL_{RF-OFF}}{PL_{RF-OFF}} \times 100\%$$

where PL represents the photoluminescence intensity of the $C_0$ line at 1571 nm, and $\Delta PL = PL_{RF-ON} - PL_{RF-OFF}$ is the difference in PL intensity with and without RF excitation.

**Data availability**

The data supporting the findings of this study are available from the corresponding author upon reasonable request.

## Acknowledgments

The authors thank Sergei Lepeshov and Søren Stobbe from Technical University of Denmark for insightful discussions and constructive feedback on the manuscript. We also acknowledge the Ion Beam Center (IBC) at Helmholtz-Zentrum Dresden-Rossendorf (HZDR) for supporting the ion implantation. This work was partially funded by the Bundesministerium für Bildung und Forschung (BMBF) under projects 13N17196 and 13N17197.

**Author contributions**

Y.B. conceived the project. S.W. performed the experiments and data analysis under the supervision of Y.B. G.P. and T.S. contributed to the theoretical description. J.Y., S.Z., and Y.B. contributed to sample fabrication and characterization. K.J., M.H., J.W.L., and S.Z. supported the project through funding and access to key instrumentation. S.W. and Y.B. wrote the manuscript with input from all authors. All authors discussed the results and contributed to the final manuscript.

**Ethics declarations**

Competing interests

The authors declare no competing interests.